\documentclass[a4paper,11pt]{article}
\pdfoutput=1
\usepackage{jheppub}
\usepackage[utf8]{inputenc}

\usepackage[vcentermath]{youngtab}
\usepackage{pdflscape}
\usepackage{tikz}
\usetikzlibrary{positioning}
\usetikzlibrary{arrows}
\usetikzlibrary{arrows.meta}
\usetikzlibrary{decorations.pathreplacing}
\usepackage{subcaption}
\usepackage{xcolor,colortbl}
\usepackage{mathtools}
\usepackage{ragged2e}
\usepackage{afterpage}
\newlength\ubwidth

\newcommand{\sslash}{\mathbin{/\mkern-6mu/}}

\tikzset{hasse/.style={circle, fill,inner sep=2pt}}
\tikzset{gauge/.style={inner sep=1mm,draw=none,fill=white,minimum size=2mm,circle, draw}}
\tikzset{gauger/.style={inner sep=1mm,draw=none,fill=red,minimum size=2mm,circle, draw}}
\tikzset{gaugeb/.style={inner sep=1mm,draw=none,fill=blue,minimum size=2mm,circle, draw}}

\usetikzlibrary{arrows,fit,decorations.pathreplacing, shapes.misc,decorations.pathmorphing}
\usepackage[capitalise]{cleveref}
\usepackage{xcolor}
\usepackage{soul}
\usepackage{colortbl}
\definecolor{Gray}{gray}{0.9}
\usepackage{multirow}
\usepackage[colorinlistoftodos]{todonotes}
\usepackage{amsthm}
\usepackage{amsmath}
\usepackage{amsthm}
\usepackage{changepage}
\usepackage{tabularx}
\usepackage{tikz}
\usepackage{multirow}
\usepackage{longtable}
\usepackage{multirow}
\usepackage{amsmath}
\usepackage{amssymb}
\usepackage{mathtools}
\usepackage{adjustbox}
\usetikzlibrary{shapes,arrows}
\usepackage[english]{babel}
\newcolumntype{P}[1]{>{\centering\arraybackslash}p{#1}}

\usepackage{color}
\usepackage{subcaption}
\usepackage{gensymb}
\usepackage{pdflscape}
\usepackage{booktabs}

\usetikzlibrary{shapes.misc}
\tikzset{gauge1/.style={draw=none,minimum size=0.6cm,fill=white,circle, draw}}
\tikzset{gauge3/.style={draw=none,minimum size=0.4cm,fill=white,circle, draw}}
\tikzset{crosses/.style={cross out, draw=black, minimum size=0.3cm, inner sep=0pt, outer sep=0pt},
cross/.default={1pt}}
\tikzset{blank/.style={draw=none,minimum size=0.4cm,fill=none,circle, draw}}
\tikzset{flavour2/.style={draw=none,minimum size=0.4cm,fill=white,regular polygon sides=4,draw}}
\tikzset{flavourBlue/.style={draw=none,minimum size=0.4cm,fill=blue,regular polygon sides=4,draw}}
\tikzset{flavourRed/.style={draw=none,minimum size=0.4cm,fill=red,regular polygon sides=4,draw}}
\tikzset{none/.style={draw=none}}
\tikzset{redgauge/.style={draw=none,minimum size=0.4cm,fill=red,circle, draw}}
\tikzset{miniU/.style={draw=none,minimum size=0.1cm,fill=red,circle, draw}}
\tikzset{smallgauge1/.style={draw=none,minimum size=0.1cm,fill=white,circle, draw}}
\tikzset{miniBlue/.style={draw=none,minimum size=0.1cm,fill=blue,circle, draw}}
\tikzset{gauge2/.style={draw=none,minimum size=0.35mm,fill=red,circle, draw}}
\tikzset{bluegauge/.style={draw=none,minimum size=0.4cm,fill=blue,circle, draw}}
\tikzset{flavour1/.style={draw=none,minimum size=0.35mm,fill=blue, regular polygon,regular polygon sides=4,draw}}
\tikzset{flavour0/.style={draw=none,minimum size=0.35mm,fill=white, regular polygon,regular polygon sides=4,draw}}
\tikzset{smalldot/.style={draw=none,minimum size=0.1mm,fill=black, circle,draw}}
\tikzset{dotsize/.style={circle,fill,inner sep=1.5pt,draw}}
\tikzset{doubleguys/.style={double, double distance = 3pt}}
\tikzset{tripleguys/.style={triple}}
\tikzset{new edge style 1/.style={dashed}}
\tikzset{thickline/.style={line width=0.06cm}}
\tikzset{brace/.style={decorate,decoration={brace,amplitude=10pt}}}
\pgfdeclarelayer{edgelayer}
\pgfdeclarelayer{nodelayer}
\usetikzlibrary{decorations.pathreplacing}
\pgfsetlayers{edgelayer,nodelayer,main}

\preprint{Imperial/TP/21/AH/02}

\title{Orthosymplectic Implosions}

\author[a]{Antoine Bourget}
\author[b]{, Andrew Dancer}
\author[a]{, Julius F. Grimminger}
\author[a]{, Amihay Hanany}
\author[c]{, Frances Kirwan}
\author[a]{and Zhenghao Zhong}
\affiliation[a]{Theoretical Physics Group, The Blackett Laboratory, Imperial College London, Prince Consort Road
London, SW7 2AZ, UK}
\affiliation[b]{Jesus College, Oxford University, OX1 3DW, UK}
\affiliation[c]{New College, Oxford University, OX1 3BN, UK}
\emailAdd{a.bourget@imperial.ac.uk}
\emailAdd{dancer@maths.ox.ac.uk}
\emailAdd{julius.grimminger17@imperial.ac.uk}
\emailAdd{a.hanany@imperial.ac.uk}
\emailAdd{kirwan@maths.ox.ac.uk}
\emailAdd{zhenghao.zhong14@imperial.ac.uk}

\abstract{We propose quivers for Coulomb branch constructions of universal implosions for orthogonal and symplectic groups, extending the work on special unitary groups in \cite{DHK}. The quivers are unitary-orthosymplectic as opposed to the purely unitary quivers in the A-type case. Where possible we check our proposals using Hilbert series techniques.}

\begin{document} 
\maketitle

\section{Introduction}

We study examples of symplectic duality, that is, duality between the Higgs and Coulomb branches of a 3-dimensional $N=4$ SUSY theory. Mathematically, this is a duality between varieties with a complex-symplectic (or even hyperK\"ahler) structure admitting a rotating circle action.

We study quivers which are candidates for the symplectic duals of universal hyperK\"ahler implosions for various groups. We briefly recall some relevant properties of these implosions here, referring the reader to \cite{guillemin2002symplectic,dancer2013implosion,dancer2016symplectic} for a more detailed description.

The universal hyperK\"ahler implosion for a complex reductive group $G = K_{\mathbb C}$ with maximal compact subgroup $K$, is supposed to be a complex-symplectic variety, that is in fact hyperK\"ahler in a suitable stratified sense. It has a complex-symplectic action of $K_{\mathbb C} \times T_{\mathbb C}$, where $T$ is the maximal torus of $K$. This action is supposed to be the complexification of a hyperK\"ahler action of the compact group $K \times T$. Moreover, the dimension of the implosion is equal to the dimension of $K_{\mathbb C} \times T_{\mathbb C}$.

HyperK\"ahler reduction of the implosion by $T$, or equivalently complex-symplectic reduction by the complex torus $T_{\mathbb C}$, will give the Kostant varieties of the Lie algebra $\frak k_{\mathbb C}$, that is, the varieties we get by fixing the values of all Ad-invariant polynomials on this Lie algebra. In particular, reduction at zero gives the nilpotent cone $\mathcal{N}_G$.  This is consistent with the dimension statements above, as each Kostant variety has dimension equal to $\dim K_{\mathbb C} - {\rm rank} \, K_{\mathbb C}$. The remaining $K_{\mathbb C}$ action on the implosion then descends to the natural action on the Kostant varieties. We can thus think of the implosion as a master space which yields the Kostant varieties on reduction by the torus action at the appropriate level.

The implosion also has a description as a Geometric Invariant Theory quotient
by the maximal unipotent subgroup $N$ of $G = K_{\mathbb C}$. Explicitly, the
implosion is the GIT quotient
\begin{equation}
(G \times {\frak n}^\circ ) \sslash N
\end{equation}
where $\frak n^{\circ}$ denotes the annihilator of $\frak n = {\rm Lie \;} N$ in the dual Lie algebra $\frak g^{*}$. We can view this quotient as the complex-symplectic quotient, in the GIT sense, of the cotangent bundle $T^*G$ by $N$. The left action of $G$ on $T^*G$ survives, but the right action is broken to a $T_{\mathbb C}$ action, giving the required action on the implosion. We recall that Kronheimer has shown the existence of a complete hyperK\"ahler metric on $T^*G$, invariant under the action of $K \times K$.

On choosing an invariant inner product on $\frak g$, we may identify ${\frak n}^\circ$ with the Borel algebra $\frak b$. Complex-symplectic reduction by $T_{\mathbb{C}} = B/N$ then fixes the Cartan component of $\frak b$ to be zero, and replaces the $N$ quotient by a $B$ quotient, yielding the GIT quotient $(G \times {\frak n}) \sslash B$. The fact that this gives the nilpotent cone is just the statement that the Springer resolution $G \times_{B} {\frak n}$ of the nilpotent cone is an affinisation map. We can view the implosion in this way as occupying an intermediate position between the cotangent bundle $T^*G$ and the nilpotent cone--the implosion is the symplectic reduction in the GIT sense of $T^*G$ by the maximal unipotent, and further reduction by the complex torus yields the nilpotent cone.

In \cite{dancer2013implosion} two of the authors and Swann produced the implosion for $K= \mathrm{SU}(n)$ as a hyperK\"ahler quiver variety. As a complex-symplectic variety the implosion is known to exist for general $K$ by work of Ginzburg-Riche \cite{Ginzburg-Riche:2015} (we note that as $N$ is nonreductive it is a nontrivial result that the quotient exists as an algebraic variety).

Implosion spaces should admit symplectic duals -- in particular as mentioned above the $\mathrm{SU}(n)$ implosion is a hyperK\"ahler quotient of a linear space at level zero, so has a $Sp(1)$ action rotating the complex structures.

In this paper we present candidates for the duals of the implosions for $K=\mathrm{SO}(2n)$ and $\mathrm{SO}(2n+1)$, extending the discussion for $\mathrm{SU}(n)$ in \cite{DHK}. One test that we will perform is to check that the dual spaces give the correct symmetry group of the original implosion. Another check is on the dimension, using Nakajima's observation that if the Higgs branch is the hyperK\"ahler quotient of a linear space $M$ by a compact group $H$, then the Coulomb branch should be birational to the quotient by the Weyl group of the cotangent bundle of the complexified dual maximal torus of $H$; hence
\begin{equation}
\dim_{\mathbb R} ( {\rm Coulomb \; branch}) =4 \; {\rm rank \;} H.
\end{equation}
We also do some checks that the hyperK\"ahler quotient of the implosion by the torus action gives the nilpotent cone, as expected.

\section{Bouquet for Orthogonal groups}{\label{implosions}}

For this section, we will focus on quivers whose Coulomb branch is the nilpotent cone of a complex semisimple group $G=K_{\mathbb C}$. We introduce the process of \textit{explosion} where a flavour node of rank $n$ is \textit{exploded} into $n$ rank 1 gauge nodes. These unbalanced nodes
generate the required Abelian symmetries in the Coulomb branch, as
required for the implosion. The number of rank 1 nodes is chosen so as to preserve the balancing condition for the remaining nodes in the original quiver, so that the $G$ symmetry in the Coulomb branch is  preserved as well, so the implosion has commuting symmetries
of $G$ and its maximal torus as required.

For unitary quivers this is well understood where we start with a $T[\mathrm{SU}(n)]$ theory with $\mathrm{SU}(n)$ flavour node. We then explode the $\mathrm{SU}(n)$ flavour node into $n$ $\mathrm{U}(1)$ gauge nodes (recall that one has to ungauge an overall $U(1)$ in the resulting unframed quiver).
\begin{equation}
 \scalebox{.800}{\begin{tikzpicture}
	\begin{pgfonlayer}{nodelayer}
		\node [style=dotsize] (5) at (-6.5, 0) {};
		\node [style=dotsize] (6) at (-6, 0) {};
		\node [style=dotsize] (7) at (-5.5, 0) {};
		\node [style=none] (15) at (-5, -0.5) {$n-1$};
		\node [style=none] (16) at (-4, -0.5) {$n$};
		\node [style=none] (17) at (-3.5, 0) {};
		\node [style=none] (18) at (-1.5, 0) {};
		\node [style=dotsize] (24) at (0.75, 0) {};
		\node [style=dotsize] (25) at (1.25, 0) {};
		\node [style=dotsize] (26) at (1.75, 0) {};
		\node [style=dotsize] (37) at (4.75, 0.5) {};
		\node [style=dotsize] (38) at (4.75, 0) {};
		\node [style=dotsize] (39) at (4.75, -0.5) {};
		\node [style=none] (42) at (5.25, 2) {1};
		\node [style=none] (43) at (5.25, 1.25) {1};
		\node [style=none] (44) at (5.25, -1.25) {1};
		\node [style=none] (45) at (5.25, -2) {1};
		\node [style=none] (46) at (5.75, 2) {};
		\node [style=none] (47) at (5.75, -2) {};
		\node [style=none] (48) at (6.5, 0) {$n$};
		\node [style=none] (49) at (-5.5, -2.25) {\scalebox{1.25}{Quiver for the nilpotent cone $\mathcal{N}_{\mathrm{SU}(n)}$}};
		\node [style=none] (50) at (-2.5, 0.75) {\scalebox{1.25}{Explosion}};
		\node [style=none] (51) at (1.75, -2.25) {\scalebox{1.25}{Exploded quiver}};
		\node [style=smallgauge1] (59) at (-7, 0) {};
		\node [style=smallgauge1] (60) at (-8, 0) {};
		\node [style=smallgauge1] (61) at (-5, 0) {};
		\node [style=flavour0] (62) at (-4, 0) {};
		\node [style=none] (63) at (-8, -0.5) {1};
		\node [style=none] (64) at (-7, -0.5) {2};
		\node [style=smallgauge1] (65) at (0.25, 0) {};
		\node [style=smallgauge1] (66) at (-0.75, 0) {};
		\node [style=none] (67) at (-0.75, -0.5) {1};
		\node [style=none] (68) at (0.25, -0.5) {2};
		\node [style=none] (69) at (3.25, -0.5) {$n-1$};
		\node [style=smallgauge1] (70) at (3.25, 0) {};
		\node [style=smallgauge1] (71) at (4.75, 2) {};
		\node [style=smallgauge1] (72) at (4.75, 1.25) {};
		\node [style=smallgauge1] (73) at (4.75, -1.25) {};
		\node [style=smallgauge1] (74) at (4.75, -2) {};
		\node [style=smallgauge1] (75) at (2.25, 0) {};
				\node [style=none] (76) at (2.25, -0.5) {$n-2$};
	\end{pgfonlayer}
	\begin{pgfonlayer}{edgelayer}
		\draw [style=->] (17.center) to (18.center);
		\draw [style=brace] (46.center) to (47.center);
		\draw (60) to (59);
		\draw (61) to (62);
		\draw (66) to (65);
		\draw (70) to (71);
		\draw (72) to (70);
		\draw (70) to (73);
		\draw (74) to (70);
		\draw (75) to (70);
	\end{pgfonlayer}
\end{tikzpicture}}
\label{SUexplosion}
\end{equation}The resulting quiver has the global symmetry changed from $\mathrm{SU}(n)$ to $\mathrm{SU}(n)\times \mathrm{U}(1)^{n-1}$. In \cite{DHK} various computational checks for the duals of these examples were made, including calculating the Hilbert series, verifying the dimension of the global symmetry group and checking Nakajima's equality, and checking that torus reduction yielded the nilpotent cone.

We now extend this to orthosymplectic quivers following the extension in the mathematical literature in \cite{dancer2016symplectic}. We adopt the usual convention in our diagrams that red nodes with label $m$ denote orthogonal groups $\mathrm{SO}(m)$\footnote{The choice of $SO$ or $O$ is important here as the Coulomb branch is sensitive to discrete factors in the gauge groups. For orthosymplectic quivers that are closures of maximal nilpotent orbits of $\mathrm{SO}(2n)$ and $\mathrm{SO}(2n+1)$, the Coulomb branch Hilbert series are computed explicitly in \cite{Cabrera:2017ucb} and contains only $SO$ and $USp$ gauge groups. } and blue nodes with label $2k$ denote symplectic groups $\mathrm{USp}(2k) = \mathrm{Sp}(k)$. We recall that a $\mathrm{USp}(2k)$ gauge node is balanced if the neighboring $\mathrm{SO}(m_j)$ nodes satisfy:
\begin{equation}
   4k = -2 + \sum\limits_{j} m_j 
\end{equation}
and an $\mathrm{SO}(m)$ node is balanced if the neighboring $\mathrm{USp}(2k_j)$ nodes satisfy:
\begin{equation}
 2m = 2 + \sum_{j} 2k_j   
\end{equation}

Note that for unframed unitary-orthosymplectic quivers, without $SO(m)$ nodes with odd $m$, the gauge group $G$ is take to be the product of the groups $G_i$ associated to the nodes $i$, modulo a diagonal $\mathbb{Z}_2$:
\begin{equation}
    G=\left(\prod_iG_i\right)/\mathbb{Z}_2 .
    \label{eq:Z2}
\end{equation}
This is not a choice of convention, but rather a necessity. Taking the gauge group to be $\prod_iG_i$ yields a different Coulomb branch which differs by a $\mathbb{Z}_2$ quotient \cite{Bourget:2020xdz}.

We start by looking at the Coulomb branch of quivers that are nilpotent cones of $\mathrm{SO}(2n)$ and $\mathrm{SO}(2n+1)$.
These are given explicitly in \cite{GaiottoWitten}. The unrefined Hilbert series of the nilpotent cone of $G$ takes a simple form: 
\begin{equation}
\mathrm{HS}_{\mathcal{N}_G} =\mathrm{PE}\left[\mathrm{dim}(G)t^2-\sum_{i=1}^r t^{2d_i}\right]
\end{equation}
where $d_i$ are the degrees of Casimir invariants of $G$ and $\mathrm{PE}$ is the plethystic exponential.  For quivers whose Coulomb branch is the nilpotent cone of $\mathrm{SO}(2n)$, the flavour node is $\mathrm{SO}(2n)$ which we explode into $n$ $\mathrm{U}(1)=\mathrm{SO}(2)$ gauge groups. The $\mathrm{USp}(2n-2)$ node thus remains balanced, as do the nodes further down the chain. The Coulomb branch Hilbert series for the exploded quivers take a more complicated form and  perturbative Hilbert series are presented in Table \ref{Dtypeimplosion}. 

Let us demonstrate this for $n=3$ which gives us the following explosion:
\begin{equation}
\scalebox{.800}{\begin{tikzpicture}
	\begin{pgfonlayer}{nodelayer}
		\node [style=redgauge] (0) at (-8, 0) {};
		\node [style=redgauge] (1) at (-6, 0) {};
		\node [style=bluegauge] (3) at (-7, 0) {};
		\node [style=bluegauge] (4) at (-5, 0) {};
		\node [style=flavourRed] (9) at (-4, 0) {};
		\node [style=none] (10) at (-8, -0.5) {2};
		\node [style=none] (11) at (-7, -0.5) {2};
		\node [style=none] (12) at (-6, -0.5) {4};
		\node [style=none] (13) at (-5, -0.5) {4};
		\node [style=none] (16) at (-4, -0.5) {6};
		\node [style=none] (17) at (-3.5, 0) {};
		\node [style=none] (18) at (-1.5, 0) {};
		\node [style=redgauge] (19) at (-1, 0) {};
		\node [style=redgauge] (20) at (1, 0) {};
		\node [style=bluegauge] (22) at (0, 0) {};
		\node [style=bluegauge] (27) at (2, 0) {};
		\node [style=none] (29) at (-1, -0.5) {2};
		\node [style=none] (30) at (0, -0.5) {2};
		\node [style=none] (31) at (1, -0.5) {4};
		\node [style=none] (34) at (2, -0.5) {4};
		\node [style=gauge3] (35) at (3, 0.75) {};
		\node [style=gauge3] (36) at (3, 0) {};
		\node [style=gauge3] (40) at (3, -0.75) {};
		\node [style=none] (42) at (3.5, 0.75) {1};
		\node [style=none] (43) at (3.5, 0) {1};
		\node [style=none] (44) at (3.5, -0.75) {1};
		\node [style=none] (49) at (-5.75, -2.25) {\scalebox{1.25}{Quiver for $\mathcal{N}_{\mathrm{SO}(6)}$}};
		\node [style=none] (50) at (-2.5, 0.75) {\scalebox{1.25}{Explosion}};
		\node [style=none] (51) at (0.5, -2.25) {\scalebox{1.25}{Exploded quiver}};
	\end{pgfonlayer}
	\begin{pgfonlayer}{edgelayer}
		\draw (0) to (3);
		\draw (3) to (1);
		\draw (1) to (4);
		\draw [style=->] (17.center) to (18.center);
		\draw (19) to (22);
		\draw (22) to (20);
		\draw (27) to (35);
		\draw (36) to (27);
		\draw (27) to (40);
		\draw (4) to (9);
		\draw (27) to (20);
	\end{pgfonlayer}
\end{tikzpicture}}
\end{equation}
Since the orthosymplectic quiver is unframed we ungauge a diagonal $\mathbb{Z}_2$, as stated in \eqref{eq:Z2}. The Coulomb branch Hilbert series can be readily computed and is found to be the same as the Coulomb branch Hilbert series of the exploded unitary quiver in \eqref{SUexplosion} for $n=4$. This is not surprising due to the isomorphism $\mathfrak{so}(6) \cong \mathfrak{su}(4)$ but nevertheless validates our approach in extending the explosion procedure to orthosymplectic quivers. 

For general $n$, we thus have:
\begin{equation}
\begin{adjustbox}{center}
   \scalebox{.800}{  \begin{tikzpicture}
	\begin{pgfonlayer}{nodelayer}
		\node [style=redgauge] (0) at (-10, 0) {};
		\node [style=redgauge] (1) at (-8, 0) {};
		\node [style=bluegauge] (3) at (-9, 0) {};
		\node [style=bluegauge] (4) at (-7, 0) {};
		\node [style=dotsize] (5) at (-6.5, 0) {};
		\node [style=dotsize] (6) at (-6, 0) {};
		\node [style=dotsize] (7) at (-5.5, 0) {};
		\node [style=bluegauge] (8) at (-5, 0) {};
		\node [style=flavourRed] (9) at (-4, 0) {};
		\node [style=none] (10) at (-10, -0.5) {2};
		\node [style=none] (11) at (-9, -0.5) {2};
		\node [style=none] (12) at (-8, -0.5) {4};
		\node [style=none] (13) at (-7, -0.5) {4};
		\node [style=none] (15) at (-5, -0.5) {$2n-2$};
		\node [style=none] (16) at (-4, -0.5) {$2n$};
		\node [style=none] (17) at (-3.5, 0) {};
		\node [style=none] (18) at (-1.5, 0) {};
		\node [style=redgauge] (19) at (-1, 0) {};
		\node [style=redgauge] (20) at (1, 0) {};
		\node [style=bluegauge] (22) at (0, 0) {};
		\node [style=bluegauge] (23) at (2, 0) {};
		\node [style=dotsize] (24) at (2.5, 0) {};
		\node [style=dotsize] (25) at (3, 0) {};
		\node [style=dotsize] (26) at (3.5, 0) {};
		\node [style=bluegauge] (27) at (4, 0) {};
		\node [style=none] (29) at (-1, -0.5) {2};
		\node [style=none] (30) at (0, -0.5) {2};
		\node [style=none] (31) at (1, -0.5) {4};
		\node [style=none] (32) at (2, -0.5) {4};
		\node [style=none] (34) at (4, -0.5) {$2n-2$};
		\node [style=gauge3] (35) at (5.5, 2) {};
		\node [style=gauge3] (36) at (5.5, 1.25) {};
		\node [style=dotsize] (37) at (5.5, 0.5) {};
		\node [style=dotsize] (38) at (5.5, 0) {};
		\node [style=dotsize] (39) at (5.5, -0.5) {};
		\node [style=gauge3] (40) at (5.5, -1.25) {};
		\node [style=gauge3] (41) at (5.5, -2) {};
		\node [style=none] (42) at (6, 2) {1};
		\node [style=none] (43) at (6, 1.25) {1};
		\node [style=none] (44) at (6, -1.25) {1};
		\node [style=none] (45) at (6, -2) {1};
		\node [style=none] (46) at (6.5, 2) {};
		\node [style=none] (47) at (6.5, -2) {};
		\node [style=none] (48) at (7.25, 0) {$n$};
		\node [style=none] (49) at (-6.5, -2.25) {\scalebox{1.25}{Quiver for $\mathcal{N}_{\mathrm{SO}(2n)}$}};
		\node [style=none] (50) at (-2.5, 0.75) {\scalebox{1.25}{Explosion}};
		\node [style=none] (51) at (2, -2.25) {\scalebox{1.25}{Exploded quiver}};
	\end{pgfonlayer}
	\begin{pgfonlayer}{edgelayer}
		\draw (0) to (3);
		\draw (3) to (1);
		\draw (1) to (4);
		\draw (8) to (9);
		\draw [style=->] (17.center) to (18.center);
		\draw (19) to (22);
		\draw (22) to (20);
		\draw (20) to (23);
		\draw (27) to (35);
		\draw (36) to (27);
		\draw (27) to (40);
		\draw (27) to (41);
		\draw [style=brace] (46.center) to (47.center);
	\end{pgfonlayer}
\end{tikzpicture}}
\end{adjustbox}
\label{soevenexplosion}
\end{equation}

The resulting global symmetry is the expected $\mathrm{SO}(2n) \times \mathrm{U}(1)^n$. The Hilbert series and plethystic logarithm for several members of this family are given in Table \ref{Dtypeimplosion}. The coefficient of $t^2$ is
\begin{equation}
2n^2 = n + 2n(2n-1)/2 = {\rm rank} \; \mathrm{SO}(2n) + \dim \mathrm{SO}(2n)\;,
\end{equation}
in agreement with the dimension of the global symmetry group. The group by which we hyperK\"ahler quotient has rank $n + 2 \sum_{i=1}^{n-1} i =n^2$, and the implosion has complex dimension $2n^2$ so real dimension $4n^2$, so Nakajima's equality is satisfied.

For quivers whose Coulomb branch is the nilpotent cone of $\mathrm{SO}(2n+1)$, the flavour node is an $\mathrm{USp}(2n)$ (the Langlands dual of $\mathrm{SO}(2n+1)$). After explosion, the flavour node explodes into $n$ $\mathrm{U}(1)$ nodes. For purposes of the balancing condition at the $\mathrm{SO}(2n)$ node, the $\mathrm{U}(1)$ nodes behave the same as $C_1 = \mathrm{USp}(2)$ (see the discussion in \cite{Bourget:2020xdz}, for example) hence the $\mathrm{SO}(2n)$ node remains balanced.

\begin{equation}
\begin{adjustbox}{center}
 \scalebox{.800}{\begin{tikzpicture}
	\begin{pgfonlayer}{nodelayer}
		\node [style=redgauge] (0) at (-10, 0) {};
		\node [style=redgauge] (1) at (-8, 0) {};
		\node [style=bluegauge] (3) at (-9, 0) {};
		\node [style=bluegauge] (4) at (-7, 0) {};
		\node [style=dotsize] (5) at (-6.5, 0) {};
		\node [style=dotsize] (6) at (-6, 0) {};
		\node [style=dotsize] (7) at (-5.5, 0) {};
		\node [style=none] (10) at (-10, -0.5) {2};
		\node [style=none] (11) at (-9, -0.5) {2};
		\node [style=none] (12) at (-8, -0.5) {4};
		\node [style=none] (13) at (-7, -0.5) {4};
		\node [style=none] (15) at (-5, -0.5) {$2n$};
		\node [style=none] (16) at (-4, -0.5) {$2n$};
		\node [style=none] (17) at (-3.5, 0) {};
		\node [style=none] (18) at (-1.5, 0) {};
		\node [style=redgauge] (19) at (-1, 0) {};
		\node [style=redgauge] (20) at (1, 0) {};
		\node [style=bluegauge] (22) at (0, 0) {};
		\node [style=bluegauge] (23) at (2, 0) {};
		\node [style=dotsize] (24) at (2.5, 0) {};
		\node [style=dotsize] (25) at (3, 0) {};
		\node [style=dotsize] (26) at (3.5, 0) {};
		\node [style=none] (29) at (-1, -0.5) {2};
		\node [style=none] (30) at (0, -0.5) {2};
		\node [style=none] (31) at (1, -0.5) {4};
		\node [style=none] (32) at (2, -0.5) {4};
		\node [style=none] (34) at (4, -0.5) {$2n$};
		\node [style=dotsize] (37) at (5.5, 0.5) {};
		\node [style=dotsize] (38) at (5.5, 0) {};
		\node [style=dotsize] (39) at (5.5, -0.5) {};
		\node [style=none] (42) at (6, 2) {1};
		\node [style=none] (43) at (6, 1.25) {1};
		\node [style=none] (44) at (6, -1.25) {1};
		\node [style=none] (45) at (6, -2) {1};
		\node [style=none] (46) at (6.5, 2) {};
		\node [style=none] (47) at (6.5, -2) {};
		\node [style=none] (48) at (7.25, 0) {$n$};
		\node [style=none] (49) at (-6.5, -2.25) {\scalebox{1.25}{Quiver for $\mathcal{N}_{\mathrm{SO}(2n+1)}$}};
		\node [style=none] (50) at (-2.5, 0.75) {\scalebox{1.25}{Explosion}};
		\node [style=none] (51) at (2, -2.25) {\scalebox{1.25}{Exploded quiver}};
		\node [style=flavourBlue] (52) at (-4, 0) {};
		\node [style=redgauge] (53) at (-5, 0) {};
		\node [style=redgauge] (54) at (4, 0) {};
		\node [style=gauge3] (55) at (5.5, 2) {};
		\node [style=gauge3] (56) at (5.5, 1.25) {};
		\node [style=gauge3] (57) at (5.5, -1.25) {};
		\node [style=gauge3] (58) at (5.5, -2) {};
	\end{pgfonlayer}
	\begin{pgfonlayer}{edgelayer}
		\draw (0) to (3);
		\draw (3) to (1);
		\draw (1) to (4);
		\draw [style=->] (17.center) to (18.center);
		\draw (19) to (22);
		\draw (22) to (20);
		\draw (20) to (23);
		\draw [style=brace] (46.center) to (47.center);
		\draw (53) to (52);
		\draw (54) to (55);
		\draw (56) to (54);
		\draw (54) to (57);
		\draw (58) to (54);
	\end{pgfonlayer}
\end{tikzpicture}}
\end{adjustbox}
\label{sooddexplosion}
\end{equation}

The Hilbert series and plethystic logarithm for several members of this family are given in Table \ref{Btypeimplosion}. 

There might be concerns about the nature of the bouquet nodes and why they are $\mathrm{U}(1)$ rather than $\mathrm{SO}(2)$ or $\mathrm{USp}(2)$. For the $\mathfrak{so}(2n)$ type quivers in (\ref{soevenexplosion}), the $\mathrm{SO}(2n)$ flavour node explodes into $n$ $\mathrm{SO}(2)$ nodes. Since $\mathrm{SO}(2) \cong \mathrm{U}(1)$, we can use either one. If we take instead the $\mathfrak{so}(2n+1)$ quivers in  (\ref{sooddexplosion}), the flavour group is $\mathrm{USp}(2n)$ so in theory we can try to explode it into $n$ $\mathrm{USp}(2)$ nodes. This computation is done in Table \ref{C1implosion}. Unfortunately, the global symmetry does not match the quivers we require for the explosion quiver family (since there are no $\mathrm{U}(1)^n$ factors). However, it does give interesting results such as for $n=4$, where there is a symmetry enhancement from $\mathrm{SO}(9)$ to $F_4$.

Another argument for the explosion of the rank $n$ flavour nodes into $n$ $\mathrm{U}(1)$ gauge nodes follows from the property that taking the hyperK\"ahler quotient of the exploded quivers by $\mathrm{U}(1)^n$ returns the quiver for the nilpotent cone, going from right to left in (\ref{soevenexplosion}) and (\ref{sooddexplosion}). This is satisfied by all our exploded quivers, hence making them good candidates for unitary-orthosymplectic quiver counterparts to the well known cases with unitary quivers (\ref{SUexplosion}).   

The $\mathrm{SO}(2n+1)$ universal implosion should have symmetry group 
$\mathrm{SO}(2n+1) \times \mathrm{U}(1)^n$, whose complex dimension is $2n(n+1)$. We see that
this appears as the $t^2$ coefficient in the Hilbert series for the 
displayed examples.

We can also check that this is consistent with Nakajima's picture. The group
by which we perform the hyperK\"ahler quotient has rank
$2 \sum_{i=1}^{n} i  = n(n+1)$, and the implosion has real dimension $4n(n+1)$, as expected. 

\subsection{HyperK\"ahler quotient}\label{hyperkahler}
Let us look at $T[G]$, the nilpotent cone of $G=\mathrm{SU}(n),\mathrm{SO}(2n),\mathrm{SO}(2n+1)$ and its explosion $\widetilde{T}[G]$. As shown in \cite{DHK}, one can recover $T[G]$ from the exploded quiver through hyperK\"ahler quotient. 

\begin{equation}
    \mathcal{C}\left(T[G]\right)=\mathcal{C}\left(\widetilde{T}[G]\right)///U(1)^{\mathrm{rank}(G)}
\end{equation}

It is well known that orthosymplectic gauge groups lack the $\mathrm{U}(1)_J$ topological symmetry that allows us to refine the Coulomb branch Hilbert series. However, we can partially refine the Coulomb branch Hilbert series by assigning fugacities $z_i$ with $i=1,\dots,n$ for the bouquet of $n$ $\mathrm{U}(1)$ gauge nodes. The fugacities carry the $\mathrm{U}(1)_J$ topological charge under each of the $\mathrm{U}(1)$ gauge nodes. On the level of the Hilbert series the hyperK\"ahler quotient with respect to these charges takes the following form:

\begin{equation}
\mathrm{HS}_{T[G]}(t)  = (1-t^2)^n \oint \prod\limits_{j=1}^{n}\frac{\mathrm{d} z_j}{2 \pi i z_j}  \mathrm{HS}_{\widetilde{T}[G]}(z_j;t)  
\end{equation}
where $G=\mathrm{SU}(n),\mathrm{SO}(2n),\mathrm{SO}(2n+1)$.

The ability to do this further justifies our choice of using a bouquet of $n$ $\mathrm{U}(1)$ nodes for $\widetilde{T}[SO(2n+1)]$ rather than a bouquet of $n$ $\mathrm{USp}(2)$ gauge nodes. This is because $\mathrm{USp}(2)\cong \mathrm{SU}(2)$ does not have a $\mathrm{U}(1)_J$ topological symmetry with which we can refine our Hilbert series.

Let us demonstrate this with $\widetilde{T}[SO(5)]$:

\begin{equation}
    \begin{split}
        &\hspace*{-1cm}(1-t^2)^2  \oint \frac{\mathrm{d}z_1 \mathrm{d}z_2}{(2 \pi i)^2 z_1 z_2}  \mathrm{HS}_{\widetilde{T}[SO(5)]}(z_1,z_2;t)=\\
        &(1-t^2)^2 \oint  \frac{\mathrm{d}z_1 \mathrm{d}z_2}{(2 \pi i)^2 z_1 z_2}  \left[1+12 t^2+ \left( \frac{4 (1+z_1^2) (1+z_2^2)}{z_1 z_2}  \right)t^3 . \right. \\
        & \left. + \left(77+\frac{5}{z_1^2}+5z_1^2+\frac{5}{z_2^2}+5z_2^2 \right) t^4 +\dots \right]\\
        &= 1 + 10 t^2 + 54 t^4 +\dots = \mathrm{HS}_{T[\mathrm{SO}(5)]}(t)
    \end{split}
\end{equation}
An exact computation with the refined Hilbert series is time consuming so we checked and matched the results to order $t^{40}$ using perturbative computations.

\afterpage{
\begin{landscape}
\begin{table}[!ht]
\small
\begin{adjustbox}{center}
	\begin{tabular}{|c|c|c|c|c|}
		\hline
Orbit &	 Quiver & \begin{tabular}{c}Global\\ Symmetry \end{tabular}& Hilbert Series  &\begin{tabular}{c} Plethystic\\ Logarithm \end{tabular}\\ 
\hline
$\mathfrak{so}(6)$ &
\begin{tabular}{c}
\centering
  \scalebox{0.75}{      
         \begin{tikzpicture}
				\tikzstyle{gauge1} = [draw=none,minimum size=0.35cm,fill=blue,circle, draw];
				\tikzstyle{gauge2} = [draw=none,minimum size=0.35cm,fill=red,circle, draw];
				\tikzstyle{flavour1} = [draw=none,minimum size=0.35cm,fill=blue, regular polygon,regular polygon sides=4,draw];
				\tikzstyle{flavour2} = [draw=none,minimum size=0.35cm,fill=red,circle, draw];
				\node (g1) [gauge2,label=below:{$2$}] {};
				\node (g2) [gauge1,right of=g1,label=below:{$2$}] {};
				\node (g3) [gauge2,right of=g2,label=below:{$4$}] {};
				\node (g4) [gauge1,right of=g3,label=below:{$4$}] {};
				\node (f1) [gauge3,right of =g4,above of=g4,label=above:{$1$}] {};
				\node (f2) [gauge3,right of =g4,label=above:{$1$}] {};
				\node (f3) [gauge3,right of =g4,below of=g4,label=above:{$1$}] {};
				\draw (g4)--(f1) (g4)--(f2) (g4)--(f3)
				(g1)--(g2) (g2)--(g3) (g3)--(g4) ;
			\end{tikzpicture}  }
\end{tabular}
& 
\begin{tabular}{c}	$\mathrm{SO}(6)$\\ $\times$ \\$ \mathrm{U}(1)^3$\end{tabular}
&
\begin{tabular}{l}
       \parbox[t]{6cm}{$ 1 + 18 t^2 + 32 t^{3} + 206 t^4 + 544 t^{5} + 1993 t^6 + 5344 t^{7} + 15531 t^8 + 39040 t^{9} + 98672 t^{10} + O\left(t^{12}\right)$ }\\
       \end{tabular}
       &
\begin{tabular}{c}
\parbox{2.5cm}{$18 t^2 + 32 t^{3} + 35 t^4 - 32 t^{5} - 305 t^6 - 672 t^{7} - 59 t^8 + 4864 t^{9} + 15662 t^{10} + O\left(t^{11}\right)$} \end{tabular}  \\ 
\hline	
$\mathfrak{so}(8)$ &
\begin{tabular}{c}	
\centering
 \scalebox{0.75}{
         \begin{tikzpicture}
				\tikzstyle{gauge1} = [draw=none,minimum size=0.35cm,fill=blue,circle, draw];
				\tikzstyle{gauge2} = [draw=none,minimum size=0.35cm,fill=red,circle, draw];
				\tikzstyle{flavour1} = [draw=none,minimum size=0.35cm,fill=blue, regular polygon,regular polygon sides=4,draw];
				\tikzstyle{flavour2} = [draw=none,minimum size=0.35cm,fill=red,circle, draw];
				\node (g1) [gauge2,label=below:{$2$}] {};
				\node (g2) [gauge1,right of=g1,label=below:{$2$}] {};
				\node (g3) [gauge2,right of=g2,label=below:{$4$}] {};
				\node (g4) [gauge1,right of=g3,label=below:{$4$}] {};
				\node (g5) [gauge2,right of=g4,label=below:{$6$}] {};
				\node (g6) [gauge1,right of=g5,label=below:{$6$}] {};	\node (f1) [gauge3,right of =g6,above of =g6, label=above:{$1$}] {};
				\node (f2) [gauge3,right of=f1 ,label=above:{$1$}] {};
				\node (f3) [gauge3,right of =g6,below of=g6 , label=below:{$1$}] {};
    			\node (f4) [gauge3,right of=f3,label=below:{$1$}] {};	\draw (g6)--(f1) (g6)--(f2) (g6)--(f3) (g6)--(f4)
				(g1)--(g2) (g2)--(g3) (g3)--(g4) (g4)--(g5) (g5)--(g6) ;
			\end{tikzpicture}  }
\end{tabular}
	& 		
\begin{tabular}{c}	$\mathrm{SO}(8)$\\ $\times$ \\$\mathrm{U}(1)^4$	\end{tabular}
&
\begin{tabular}{c} \parbox{6cm}{
$1+32 t^2+527 t^4+6144 t^6+57782 t^8+466656 t^{10}+3339801 t^{12}+21550576 t^{14}+126756294 t^{16}+685145416 t^{18}+3426352669 t^{20}+ O\left(t^{22}\right)$}
\end{tabular}
	&
\begin{tabular}{c}
\parbox[t]{2.5cm}{$32 t^2 - t^4 + 192 t^6 - 194 t^8 + 672 t^{10} +O\left(t^{11}\right)$}
\end{tabular} \\ 
\hline
 $ \mathfrak{so}(10)$ 
 &   
\begin{tabular}{c}	
\centering
\scalebox{0.75}{
\begin{tikzpicture}{center}
				\tikzstyle{gauge1} = [draw=none,minimum size=0.35cm,fill=blue,circle, draw];
				\tikzstyle{gauge2} = [draw=none,minimum size=0.35cm,fill=red,circle, draw];
				\tikzstyle{flavour1} = [draw=none,minimum size=0.35cm,fill=blue, regular polygon,regular polygon sides=4,draw];
				\tikzstyle{flavour2} = [draw=none,minimum size=0.35cm,fill=red,circle, draw];
				\node (g1) [gauge2,label=below:{$2$}] {};
				\node (g2) [gauge1,right of=g1,label=below:{$2$}] {};
				\node (g3) [gauge2,right of=g2,label=below:{$4$}] {};
				\node (g4) [gauge1,right of=g3,label=below:{$4$}] {};
				\node (g5) [gauge2,right of=g4,label=below:{$6$}] {};
				\node (g6) [gauge1,right of=g5,label=below:{$6$}] {};
				\node (g7) [gauge2,right of=g6,label=below:{$8$}] {};
				\node (g8) [gauge1,right of=g7,label=below:{$8$}] {};
				\node (f2) [gauge3,right of =g8,above of=g8, label=above:{$1$}] {};
				\node (f1) [gauge3,right of=f2,label=above:{$1$}] {};  \node (f3) [gauge3,right of =g8,below of=g8,label=below:{$1$}] {};
				\node (f4) [gauge3,right of=f3,label=below:{$1$}] {};	\node (f5) [gauge3,right of=g8,label=right:{$1$}] {};	\draw  (g8)--(f2) (g8)--(f1) (g8)--(f3)(g8)--(f4)(g8)--(f5)
				(g1)--(g2) (g2)--(g3) (g3)--(g4) (g4)--(g5) (g5)--(g6) (g6)--(g7) (g7)--(g8) ;
				\end{tikzpicture}}
\end{tabular}
	& 		
\begin{tabular}{c}	$\mathrm{SO}(10)$\\ $\times$ \\$\mathrm{U}(1)^5$	\end{tabular}
&
\begin{tabular}{c}
\parbox[t]{6cm}{$1+50 t^2+1274 t^4+22050 t^6+291649 t^8+3145771 t^{10}+28843013 t^{12}+231437604 t^{14}+1660782225 t^{16}+10838595181 t^{18}+65196668471 t^{20}+O\left(t^{22}\right)$}\end{tabular}
	&
\begin{tabular}{c}
\parbox[t]{2.5cm}{$50 t^2-t^4+99 t^8+411 t^{10} +O\left(t^{11}\right)$}
\end{tabular} \\ 
\hline
$ \mathfrak{so}(12)$ &   
\begin{tabular}{c}	
\centering
\scalebox{0.6}{
\begin{tikzpicture}{center}
				\tikzstyle{gauge1} = [draw=none,minimum size=0.35cm,fill=blue,circle, draw];
				\tikzstyle{gauge2} = [draw=none,minimum size=0.35cm,fill=red,circle, draw];
				\tikzstyle{flavour1} = [draw=none,minimum size=0.35cm,fill=blue, regular polygon,regular polygon sides=4,draw];
				\tikzstyle{flavour2} =  = [draw=none,minimum size=0.35cm,fill=red,circle, draw];
				\node (g1) [gauge2,label=below:{$2$}] {};
				\node (g2) [gauge1,right of=g1,label=below:{$2$}] {};
				\node (g3) [gauge2,right of=g2,label=below:{$4$}] {};
				\node (g4) [gauge1,right of=g3,label=below:{$4$}] {};
				\node (g5) [gauge2,right of=g4,label=below:{$6$}] {};
				\node (g6) [gauge1,right of=g5,label=below:{$6$}] {};
				\node (g7) [gauge2,right of=g6,label=below:{$8$}] {};
				\node (g8) [gauge1,right of=g7,label=below:{$8$}] {};
				\node (g9) [gauge2,right of=g8,label=below:{$10$}] {};
				\node (g10) [gauge1,right of=g9,label=below:{$10$}] {};	\node (f2) [gauge3,right of =g10,above of=g10, label=above:{$1$}] {};
				\node (f1) [gauge3,right of=f2,label=above:{$1$}] {};  \node (f3) [gauge3,right of =g10,below of=g10,label=below:{$1$}] {};
				\node (f4) [gauge3,right of=f3,label=below:{$1$}] {};	\node (f5) [gauge3,right of=f4,label=below:{$1$}] {};	\node (f6) [gauge3,right of=f1,label=above:{$1$}] {};	\draw  (g10)--(f2) (g10)--(f1) (g10)--(f3)(g10)--(f4)(g10)--(f5) (g10)--(f6)
				(g1)--(g2) (g2)--(g3) (g3)--(g4) (g4)--(g5) (g5)--(g6) (g6)--(g7) (g7)--(g8) (g8)--(g9) (g9)--(g10);
				\end{tikzpicture}}
\end{tabular}
	& 		
\begin{tabular}{c}	$\mathrm{SO}(12)$\\ $\times$ \\$\mathrm{U}(1)^6$	\end{tabular}
&
\begin{tabular}{c} 
\parbox[t]{6cm}{$1 + 72 t^2 + 2627 t^4 + 64752 t^6 + 1212821 t^8 + 18410088 t^{10} + 235885925 t^{12} + 2623730304 t^{14} + 2048 t^{15} + 25859417578 t^{16} + 145408 t^{17} + 229405181384 t^{18} + 5232640 t^{19} + 1854541984877 t^{20} + 127229952 t^{21}+O \left(t^{22}\right)$}
\end{tabular}
&
\begin{tabular}{c}
\parbox[t]{2.5cm}{$72 t^2 - t^4 + - t^8 + 144 t^{10} + O\left(t^{11}\right)$}
\end{tabular} \\ 
\hline
$ \mathfrak{so}(14)$ &   
\begin{tabular}{c}	
\centering
\scalebox{0.5}{
\begin{tikzpicture}{center}
				\tikzstyle{gauge1} = [draw=none,minimum size=0.35cm,fill=blue,circle, draw];
				\tikzstyle{gauge2} = [draw=none,minimum size=0.35cm,fill=red,circle, draw];
				\tikzstyle{flavour1} =  = [draw=none,minimum size=0.35cm,fill=blue,circle, draw];
				\node (g1) [gauge2,label=below:{$2$}] {};
				\node (g2) [gauge1,right of=g1,label=below:{$2$}] {};
				\node (g3) [gauge2,right of=g2,label=below:{$4$}] {};
				\node (g4) [gauge1,right of=g3,label=below:{$4$}] {};
				\node (g5) [gauge2,right of=g4,label=below:{$6$}] {};
				\node (g6) [gauge1,right of=g5,label=below:{$6$}] {};
				\node (g7) [gauge2,right of=g6,label=below:{$8$}] {};
				\node (g8) [gauge1,right of=g7,label=below:{$8$}] {};
				\node (g9) [gauge2,right of=g8,label=below:{$10$}] {};
				\node (g10) [gauge1,right of=g9,label=below:{$10$}] {};
				\node (g11) [gauge2,right of=g10,label=below:{$12$}] {};
				\node (g12) [gauge1,right of=g11,label=below:{$12$}] {};
				\node (f2) [gauge3,right of =g12,above of=g12, label=above:{$1$}] {};
				\node (f1) [gauge3,right of=f2,label=above:{$1$}] {};
				\node (f3) [gauge3,right of =g12,below of=g12,label=below:{$1$}] {};
				\node (f4) [gauge3,right of=f3,label=below:{$1$}] {};
				\node (f5) [gauge3,right of=f4,label=below:{$1$}] {};	\node (f6) [gauge3,right of=f1,label=above:{$1$}] {};
				\node (f7) [gauge3,right of=g12,label=right:{$1$}] {};	\draw  (g12)--(f2) (g12)--(f1) (g12)--(f3)(g12)--(f4)(g12)--(f5) (g12)--(f6) (g12)--(f7)
				(g1)--(g2) (g2)--(g3) (g3)--(g4) (g4)--(g5) (g5)--(g6) (g6)--(g7) (g7)--(g8) (g8)--(g9) (g9)--(g10) (g10)--(g11) (g11)--(g12);
				\end{tikzpicture}}
\end{tabular}
	& 		
\begin{tabular}{c}	$\mathrm{SO}(14)$\\ $\times$ \\$\mathrm{U}(1)^7$	\end{tabular}
&	
\begin{tabular}{c}  
\parbox[t]{6cm}{$1+98 t^2+4850 t^4+161602 t^6+4078073 t^8+83129872 t^{10}+1425752755 t^{12}+O \left(t^{14}\right)$}
\end{tabular}
&
\begin{tabular}{c}
\parbox[t]{2.5cm}{$98 t^2 - t^4 -t^8 \left(t^{11}\right)$}
\end{tabular} \\ 
\hline
\end{tabular}
\end{adjustbox}
\caption{Quivers whose Coulomb branches are the nilpotent cones of $\mathrm{SO}(2n)$ for $n=3,4,5,6,7$ are fully `exploded'. The $\mathrm{SO}(2n)$ flavour node turns into $n$ $\mathrm{U}(1)$ gauge nodes.  The Hilbert series and the plethystic logarithm are provided.
}
\label{Dtypeimplosion}

\end{table}
\begin{table}[!ht]

\begin{adjustbox}{center}
\scalebox{0.75}{	\begin{tabular}{|l|l|l|l|l|}
		\hline
Orbit &	 Quiver &Global Symmetry&Hilbert Series  &Plethystic Logarithm \\ \hline
$\mathfrak{so}(5)$ &

\begin{tabular}{c}
\centering
         
    \begin{tikzpicture}
				\tikzstyle{gauge1} = [draw=none,minimum size=0.35cm,fill=blue,circle, draw];
				\tikzstyle{gauge2} = [draw=none,minimum size=0.35cm,fill=red,circle, draw];
            	\tikzstyle{flavour1}  = [draw=none,minimum size=0.35cm,fill=blue,circle, draw];
				\tikzstyle{flavour2} = [draw=none,minimum size=0.35cm,fill=red, regular polygon,regular polygon sides=4,draw];
				\node (g5) [gauge2,right of=g4,label=below:{$2$}] {};
				\node (g6) [gauge1,right of=g5,label=below:{$2$}] {};
				\node (g7) [gauge2,right of=g6,label=below:{$4$}] {};

				\node (f2) [gauge3,right of =g7,above of=g7, label=above:{$1$}] {};
				\node (f3) [gauge3,right of =g7,below of=g7,label=below:{$1$}] {};
				\draw (g7)--(f2) (g7)--(f3)   (g5)--(g6)  (g6)--(g7);
				\end{tikzpicture}  
\end{tabular}
& 		
\centering	\begin{tabular}{c}	$\mathrm{SO}(5) \times \mathrm{U}(1)^2$\end{tabular}	&
\begin{tabular}{l}
       \parbox[t]{5cm}{{$1 + 12 t^2 + 16 t^3 + 97 t^4 + 176 t^5 + 612 t^6 + 1168 t^7 + 
 3054 t^8 + 5728 t^9 + 12640 t^{10} + 22768 t^{11} + 44842 t^{12} + 
 77312 t^{13} + 140220 t^{14} + 231936 t^{15} + 394955 t^{16} + 629232 t^{17} + 
 1019236 t^{18} + 1570512 t^{19} + 2442031 t^{20}+O\left(t^{22}\right)$ } }\\
       \end{tabular}& \begin{tabular}{c}\parbox{2.5cm}{$ 12 t^2 + 16 t^3 + 19 t^4 - 16 t^5 - 116 t^6 - 192 t^7 + 33 t^8 + 
 1152 t^9 + 2764 t^{10}+O\left(t^{12}\right)   $} \end{tabular}  \\  \hline	
   $\mathfrak{so}(7)$ &    
\begin{tabular}{c}
\centering
    \begin{tikzpicture}
				\tikzstyle{gauge1} = [draw=none,minimum size=0.35cm,fill=blue,circle, draw];
				\tikzstyle{gauge2} = [draw=none,minimum size=0.35cm,fill=red,circle, draw];
            	\tikzstyle{flavour1}  = [draw=none,minimum size=0.35cm,fill=blue,circle, draw];
				\tikzstyle{flavour2} = [draw=none,minimum size=0.35cm,fill=red, regular polygon,regular polygon sides=4,draw];
				\node (g3) [gauge2,right of=g2,label=below:{$2$}] {};
				\node (g4) [gauge1,right of=g3,label=below:{$2$}] {};
				\node (g5) [gauge2,right of=g4,label=below:{$4$}] {};
				\node (g6) [gauge1,right of=g5,label=below:{$4$}] {};
				\node (g7) [gauge2,right of=g6,label=below:{$6$}] {};
				\node (f2) [gauge3,right of =g7,above of=g7, label=above:{$1$}] {};
				\node (f1) [gauge3,right of=g7,label=right:{$1$}] {};                
				\node (f3) [gauge3,right of =g7,below of=g7,label=below:{$1$}] {};
				\draw (g7)--(f1) (g7)--(f2) (g7)--(f3) (g3)--(g4) (g4)--(g5) (g5)--(g6)  (g6)--(g7);
				\end{tikzpicture}  
\end{tabular}
& 		\centering	\begin{tabular}{c}		$\mathrm{SO}(7) \times \mathrm{U}(1)^3$\end{tabular}	&\begin{tabular}{l}
       \parbox[t]{5cm}{{$1 + 24 t^2 + 299 t^4 + 2682 t^6 + 19687 t^8 + 125058 t^{10} + 
 705840 t^{12} + 3592368 t^{14} + 16656892 t^{16} + 70957310 t^{18} + 
 279781688 t^{20}+O\left(t^{22}\right)$ } }\\
       \end{tabular}& \begin{tabular}{c}\parbox{2.5cm}{$ 24 t^2 - t^4 + 106 t^6 - 107 t^8 + 252 t^{10} +O\left(t^{12}\right)   $} \end{tabular}  \\  \hline	
       $\mathfrak{so}(9)$ &
\begin{tabular}{c}
\centering
         
    \begin{tikzpicture}
				\tikzstyle{gauge1} = [draw=none,minimum size=0.35cm,fill=blue,circle, draw];
				\tikzstyle{gauge2} = [draw=none,minimum size=0.35cm,fill=red,circle, draw];
            	\tikzstyle{flavour1}  = [draw=none,minimum size=0.35cm,fill=blue,circle, draw];
				\tikzstyle{flavour2} = [draw=none,minimum size=0.35cm,fill=red, regular polygon,regular polygon sides=4,draw];
				\node (g1) [gauge2,label=below:{$2$}] {};
				\node (g2) [gauge1,right of=g1,label=below:{$2$}] {};
				\node (g3) [gauge2,right of=g2,label=below:{$4$}] {};
				\node (g4) [gauge1,right of=g3,label=below:{$4$}] {};
				\node (g5) [gauge2,right of=g4,label=below:{$6$}] {};
				\node (g6) [gauge1,right of=g5,label=below:{$6$}] {};
				\node (g7) [gauge2,right of=g6,label=below:{$8$}] {};
				\node (f2) [gauge3,right of =g7,above of=g7, label=above:{$1$}] {};
				\node (f1) [gauge3,right of=f2,label=above:{$1$}] {};                
				\node (f3) [gauge3,right of =g7,below of=g7,label=below:{$1$}] {};
				\node (f4) [gauge3,right of=f3,label=below:{$1$}] {};
				\draw (g7)--(f1) (g7)--(f2) (g7)--(f3) (g7)--(f4) 
				(g1)--(g2) (g2)--(g3) (g3)--(g4) (g4)--(g5) (g5)--(g6)  (g6)--(g7);
				\end{tikzpicture}  
\end{tabular}
& 		
\centering	\begin{tabular}{c}		$\mathrm{SO}(9) \times \mathrm{U}(1)^4$\end{tabular}	&
\begin{tabular}{l}
       \parbox[t]{5cm}{{$1 + 40 t^2 + 819 t^4 + 11440 t^6 + 122661 t^8 + 1077552 t^{ 10} + 8086902 t^{12} + 53392192 t^{14} 
 316944489 t^{16}  + 1720172104 t^{18} + 
 8648640839 t^{20}+O\left(t^{22}\right)$ } }\\

       \end{tabular}& \begin{tabular}{c}\parbox{2.5cm}{$ 40 t^2 - t^4 + 71 t^8 + 184 t^{10}  +O\left(t^{12}\right)   $} \end{tabular}  \\  \hline	

       $\mathfrak{so}(11)$&
\begin{tabular}{c}	
\centering
 
         \begin{tikzpicture}
				\tikzstyle{gauge1} = [draw=none,minimum size=0.35cm,fill=blue,circle, draw];
				\tikzstyle{gauge2} = [draw=none,minimum size=0.35cm,fill=red,circle, draw];
            	\tikzstyle{flavour1}  = [draw=none,minimum size=0.35cm,fill=blue,circle, draw];
				\tikzstyle{flavour2} = [draw=none,minimum size=0.35cm,fill=red, regular polygon,regular polygon sides=4,draw];
				\node (g1) [gauge2,label=below:{$2$}] {};
				\node (g2) [gauge1,right of=g1,label=below:{$2$}] {};
				\node (g3) [gauge2,right of=g2,label=below:{$4$}] {};
				\node (g4) [gauge1,right of=g3,label=below:{$4$}] {};
				\node (g5) [gauge2,right of=g4,label=below:{$6$}] {};
				\node (g6) [gauge1,right of=g5,label=below:{$6$}] {};
				\node (g7) [gauge2,right of=g6,label=below:{$8$}] {};
				\node (g8) [gauge1,right of=g7,label=below:{$8$}] {};
				\node (g9) [gauge2,right of=g8,label=below:{$10$}] {};
				\node (f2) [gauge3,right of =g9,above of=g9, label=above:{$1$}] {};
				\node (f1) [gauge3,right of=f2,label=above:{$1$}] {};                
				\node (f3) [gauge3,right of =g9,below of=g9,label=below:{$1$}] {};
				\node (f4) [gauge3,right of=f3,label=below:{$1$}] {};
			\node (f5) [gauge3,right of=g9,label=right:{$1$}] {};	
				\draw (g9)--(f5) (g9)--(f1) (g9)--(f2) (g9)--(f3) (g9)--(f4) 
				(g1)--(g2) (g2)--(g3) (g3)--(g4) (g4)--(g5) (g5)--(g6)  (g6)--(g7) (g7)--(g8) (g8)--(g9);
				\end{tikzpicture}  
\end{tabular}
	& 		
	\centering	\begin{tabular}{c}	$\mathrm{SO}(11)\times \mathrm{U}(1)^5$	\end{tabular}&	\begin{tabular}{c}  \parbox[t]{5cm}{$1 + 60 t^2 + 1829 t^4 + 37760 t^6 + 593834 t^8 + 7586742 t^{10} + 
 82007875 t^{12} + 771321698 t^{14} + 1024 t^{15} + 6443119425 t^{16} + 
 60416 t^{17} + 48554821508 t^{18} + 1811456 t^{19} + 334196827069 t^{20}+36792320 t^{21}+O\left(t^{22}\right)$}\end{tabular}
	&\begin{tabular}{c}\parbox[t]{2.5cm}{$60 t^2 - t^4 - t^8 + 110 t^{10} - 111 t^{12 }+ 1024 t^{15} +O\left(t^{16}\right)$}\end{tabular} \\ \hline
 	 $ \mathfrak{so}(13)$ &   
\begin{tabular}{c}	
\centering
\begin{tikzpicture}
				\tikzstyle{gauge1} = [draw=none,minimum size=0.35cm,fill=blue,circle, draw];
				\tikzstyle{gauge2} = [draw=none,minimum size=0.35cm,fill=red,circle, draw];
            	\tikzstyle{flavour1}  = [draw=none,minimum size=0.35cm,fill=blue,circle, draw];
				\tikzstyle{flavour2} = [draw=none,minimum size=0.35cm,fill=red, regular polygon,regular polygon sides=4,draw];
				\node (g1) [gauge2,label=below:{$2$}] {};
				\node (g2) [gauge1,right of=g1,label=below:{$2$}] {};
				\node (g3) [gauge2,right of=g2,label=below:{$4$}] {};
				\node (g4) [gauge1,right of=g3,label=below:{$4$}] {};
				\node (g5) [gauge2,right of=g4,label=below:{$6$}] {};
				\node (g6) [gauge1,right of=g5,label=below:{$6$}] {};
				\node (g7) [gauge2,right of=g6,label=below:{$8$}] {};
				\node (g8) [gauge1,right of=g7,label=below:{$8$}] {};
				\node (g9) [gauge2,right of=g8,label=below:{$10$}] {};
		    	\node (g10) [gauge1,right of=g9,label=below:{$10$}] {};	
		    	\node (g11) [gauge2,right of=g10,label=below:{$12$}] {};	
				\node (f2) [gauge3,right of =g11,above of=g11, label=above:{$1$}] {};
				\node (f1) [gauge3,right of=f2,label=above:{$1$}] {};                
				\node (f3) [gauge3,right of =g11,below of=g11,label=below:{$1$}] {};
				\node (f4) [gauge3,right of=f3,label=below:{$1$}] {};
			\node (f5) [gauge3,right of=f4,label=below:{$1$}] {};
				\node (f6) [gauge3,right of=f1,label=above:{$1$}] {};
				\draw (g11)--(f5) (g11)--(f1) (g11)--(f2) (g11)--(f3) (g11)--(f4) 
				(g1)--(g2) (g2)--(g3) (g3)--(g4) (g4)--(g5) (g5)--(g6)  (g6)--(g7) (g7)--(g8) (g8)--(g9) (g9)--(g10) (g10)--(g11) (g11)--(f6);
				\end{tikzpicture}  
\end{tabular}
	& 		
	\centering	\begin{tabular}{c}	$\mathrm{SO}(13)\times \mathrm{U}(1)^6$	\end{tabular}&	\begin{tabular}{c}  \parbox[t]{5cm}{$1 + 84 t^2 + 3569 t^4 + 102256 t^6 + 2222324 t^8 + 39073328 t^{10} + 
 578877666 t^{12} + 7432109337 t^{14} + 84404062467 t^{16} + 
 861258507989 t^{18} + 7994156173400 t^{20}+4096t^{21}+O\left(t^{22}\right)$}\end{tabular}
	&\begin{tabular}{c}\parbox[t]{2.5cm}{$84 t^2 - t^4 - t^8 + 142 t^{12} - 143 t^{14} - t^{16} - t^{20} + 4096 t^{21} +O\left(t^{22}\right)$}\end{tabular} \\ \hline
	\end{tabular}}
\end{adjustbox}
\caption{Quivers whose Coulomb branches are the nilpotent cones of $\mathrm{SO}(2n+1)$ for $n=4,5,6$ are fully `exploded'. The $\mathrm{USp}(2n)$ flavour node turns into $n$ $\mathrm{U}(1)$ gauge nodes. The Hilbert series and the plethystic logarithm are provided.}
\label{Btypeimplosion}
\end{table}
 \end{landscape}}

\paragraph{Integer and half-integer magnetic lattice}\label{Hilbertseries}
In \cite{Bourget:2020xdz}, it has been pointed out for unframed orthosymplectic quivers made of $\mathrm{SO}(2n)$, $\mathrm{USp}(2k)$ and $\mathrm{U}(m)$ gauge groups, one can ungauge an overall diagonal $\mathbb{Z}_2$. The effect of ungauging this $\mathbb{Z}_2$ translates to changing the magnetic lattice of the gauge groups to include half-integer magnetic charges. Therefore, we can write the Hilbert series as $\mathrm{HS}=\mathrm{HS}_{\mathbb{Z}}+\mathrm{HS}_{\mathbb{Z}+\frac{1}{2}}$ where $\mathrm{HS}_{\mathbb{Z}}$ comes from integer magnetic charge contributions and $\mathrm{HS}_{\mathbb{Z}+\frac{1}{2}}$ from half-integer magnetic charge contributions. One thing to notice is that all the operators from $\mathrm{HS}_{\mathbb{Z}+\frac{1}{2}}$ carry non-trivial $\mathrm{U}(1)_J$ charges from \emph{every} $\mathrm{U}(1)$ node in the bouquet. Therefore, the contribution from $\mathrm{HS}_{\mathbb{Z}+\frac{1}{2}}$ is zero after taking a hyperK\"ahler quotient. This result is expected because taking a hyperK\"ahler quotient over a $\mathrm{U}(1)$ gauge node is equivalent to turning it into a $\mathrm{U}(1)$ flavor node. Therefore, even if we quotient over one of the $\mathrm{U}(1)$ nodes in the bouquet, it changes the exploded quiver from an unframed orthosymplectic quiver to a framed orthosymplectic quiver. And for a framed orthosymplectic quiver,  the magnetic lattice only contains integer magnetic charges and the Hilbert series is $\mathrm{HS}_{\mathbb{Z}}$.

\section{Bouquet for Symplectic groups}
Above, bouquets for $G=\mathrm{SU}(n), \mathrm{SO}(2n), \mathrm{SO}(2n+1)$ are investigated. The reason we haven't investigated $G=\mathrm{USp}(2n)$ is because the orthosymplectic quivers are all `bad'. In particular, some of the gauge nodes have negative balance. Therefore, the monopole formula diverges and we cannot test any of our conjectures through explicit computation. However, orthosymplectic quivers whose Coulomb branch are conjectured to be closures of $G=\mathrm{USp}(2n)$ nilpotent orbits had been studied through other means, in particular through brane configurations in \cite{Feng:2000eq,GaiottoWitten,Cabrera:2017njm}. With this, one can construct the quiver for $T[\mathrm{USp}(2n)]$ where the $\mathrm{USp}(2n)$ global symmetry can be read off following a new set of balancing conditions. Since the quivers are obtained from brane configurations which are not sensitive to the difference between $\mathrm{SO}$ and $\mathrm{O}$ gauge groups, we will stick with their algebras $\mathfrak{so}$ in this section. For $\mathfrak{usp}(2k)$, we need $\mathfrak{so}(m_j)$ neighboring nodes such that $4k=\sum_j mj$. For $\mathfrak{so}(2m+1)$, we need $\mathfrak{usp}(2k_j)$ neighboring nodes such that $4m+2=\sum_j 2k_j$. For $2n$ nodes arranged linearly satisfying these balancing conditions will give a $\mathrm{USp}(2n)$. For $T[\mathrm{USp}(2n)]$, the flavor node is $\mathfrak{so}(2n+1)$ and the gauge node it is connected to is $\mathfrak{usp}(2n)$. A very natural partition of the bouquet is into $\mathfrak{so}(1)$ and $n$ $\mathfrak{u}(1)$s. Here, we know that the global symmetry for $\mathrm{U}(1)$ with $n$ flavors where $n >2$ is $\mathrm{U}(1)$. Therefore, the global symmetry of the following explosion based on the balance of the gauge nodes give the expected $G_{\mathrm{global}}=\mathrm{USp}(2n)\times \mathrm{U}(1)^n$. Note that the sum of the ranks of the 
groups at the gauge nodes is $n + \sum_{i=1}^{n} i +
\sum_{i=1}^{n-1} i = n(n+1)$, as in the $SO(2n+1)$
case discussed in the previous section.
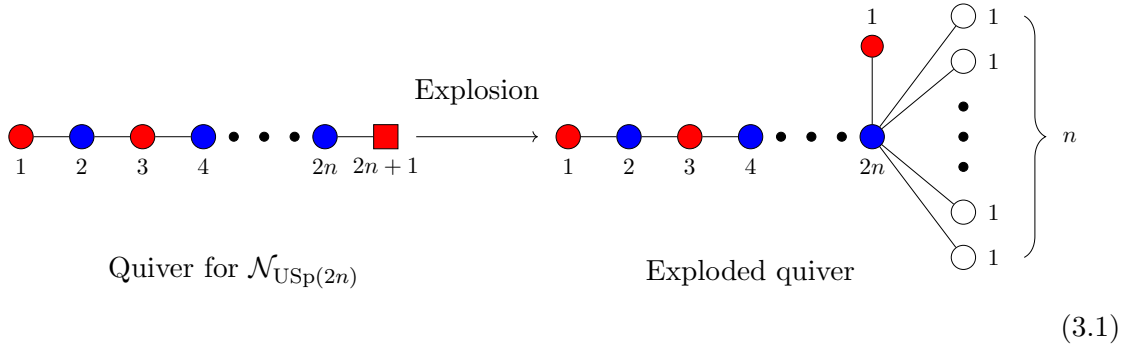
\begin{equation}
\begin{adjustbox}{center}
 \scalebox{.800}{\begin{tikzpicture}
	\begin{pgfonlayer}{nodelayer}
		\node [style=redgauge] (0) at (-10, 0) {};
		\node [style=redgauge] (1) at (-8, 0) {};
		\node [style=bluegauge] (3) at (-9, 0) {};
		\node [style=bluegauge] (4) at (-7, 0) {};
		\node [style=dotsize] (5) at (-6.5, 0) {};
		\node [style=dotsize] (6) at (-6, 0) {};
		\node [style=dotsize] (7) at (-5.5, 0) {};
		\node [style=bluegauge] (8) at (-5, 0) {};
		\node [style=flavourRed] (9) at (-4, 0) {};
		\node [style=none] (10) at (-10, -0.5) {1};
		\node [style=none] (11) at (-9, -0.5) {2};
		\node [style=none] (12) at (-8, -0.5) {3};
		\node [style=none] (13) at (-7, -0.5) {4};
		\node [style=none] (15) at (-5, -0.5) {$2n$};
		\node [style=none] (16) at (-4, -0.5) {$2n+1$};
		\node [style=none] (17) at (-3.5, 0) {};
		\node [style=none] (18) at (-1.5, 0) {};
		\node [style=redgauge] (19) at (-1, 0) {};
		\node [style=redgauge] (20) at (1, 0) {};
		\node [style=bluegauge] (22) at (0, 0) {};
		\node [style=bluegauge] (23) at (2, 0) {};
		\node [style=dotsize] (24) at (2.5, 0) {};
		\node [style=dotsize] (25) at (3, 0) {};
		\node [style=dotsize] (26) at (3.5, 0) {};
		\node [style=bluegauge] (27) at (4, 0) {};
		\node [style=none] (29) at (-1, -0.5) {1};
		\node [style=none] (30) at (0, -0.5) {2};
		\node [style=none] (31) at (1, -0.5) {3};
		\node [style=none] (32) at (2, -0.5) {4};
		\node [style=none] (34) at (4, -0.5) {$2n$};
		\node [style=gauge3] (35) at (5.5, 2) {};
		\node [style=gauge3] (36) at (5.5, 1.25) {};
		\node [style=dotsize] (37) at (5.5, 0.5) {};
		\node [style=dotsize] (38) at (5.5, 0) {};
		\node [style=dotsize] (39) at (5.5, -0.5) {};
		\node [style=gauge3] (40) at (5.5, -1.25) {};
		\node [style=gauge3] (41) at (5.5, -2) {};
		\node [style=none] (42) at (6, 2) {1};
		\node [style=none] (43) at (6, 1.25) {1};
		\node [style=none] (44) at (6, -1.25) {1};
		\node [style=none] (45) at (6, -2) {1};
		\node [style=none] (46) at (6.5, 2) {};
		\node [style=none] (47) at (6.5, -2) {};
		\node [style=none] (48) at (7.25, 0) {$n$};
		\node [style=none] (49) at (-6.5, -2.25) {\scalebox{1.25}{Quiver for $\mathcal{N}_{\mathrm{USp}(2n)}$}};
		\node [style=none] (50) at (-2.5, 0.75) {\scalebox{1.25}{Explosion}};
		\node [style=none] (51) at (2, -2.25) {\scalebox{1.25}{Exploded quiver}};
		\node [style=none] (53) at (4, 2) {1};
		\node [style=miniU] (54) at (4, 1.5) {};
	\end{pgfonlayer}
	\begin{pgfonlayer}{edgelayer}
		\draw (0) to (3);
		\draw (3) to (1);
		\draw (1) to (4);
		\draw (8) to (9);
		\draw [style=->] (17.center) to (18.center);
		\draw (19) to (22);
		\draw (22) to (20);
		\draw (20) to (23);
		\draw (27) to (35);
		\draw (36) to (27);
		\draw (27) to (40);
		\draw (27) to (41);
		\draw [style=brace] (46.center) to (47.center);
		\draw (54) to (27);
	\end{pgfonlayer}
\end{tikzpicture}}
\end{adjustbox}
\label{implodeSO(2n+1)}
\end{equation}

Another natural partition is into a $\mathfrak{so}(3)$ and $n-1$ $\mathfrak{u}(1)$s. The exploded quiver takes the following form:
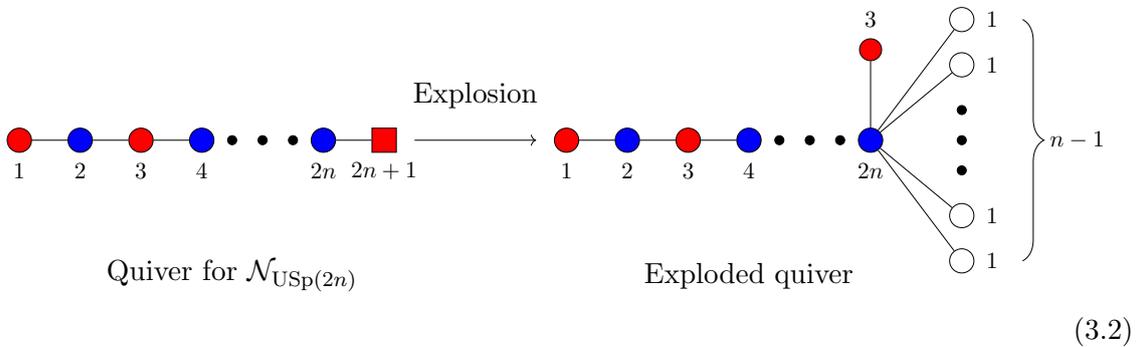
\begin{equation}
\begin{adjustbox}{center}
 \scalebox{.800}{\begin{tikzpicture}
	\begin{pgfonlayer}{nodelayer}
		\node [style=redgauge] (0) at (-10, 0) {};
		\node [style=redgauge] (1) at (-8, 0) {};
		\node [style=bluegauge] (3) at (-9, 0) {};
		\node [style=bluegauge] (4) at (-7, 0) {};
		\node [style=dotsize] (5) at (-6.5, 0) {};
		\node [style=dotsize] (6) at (-6, 0) {};
		\node [style=dotsize] (7) at (-5.5, 0) {};
		\node [style=bluegauge] (8) at (-5, 0) {};
		\node [style=flavourRed] (9) at (-4, 0) {};
		\node [style=none] (10) at (-10, -0.5) {1};
		\node [style=none] (11) at (-9, -0.5) {2};
		\node [style=none] (12) at (-8, -0.5) {3};
		\node [style=none] (13) at (-7, -0.5) {4};
		\node [style=none] (15) at (-5, -0.5) {$2n$};
		\node [style=none] (16) at (-4, -0.5) {$2n+1$};
		\node [style=none] (17) at (-3.5, 0) {};
		\node [style=none] (18) at (-1.5, 0) {};
		\node [style=redgauge] (19) at (-1, 0) {};
		\node [style=redgauge] (20) at (1, 0) {};
		\node [style=bluegauge] (22) at (0, 0) {};
		\node [style=bluegauge] (23) at (2, 0) {};
		\node [style=dotsize] (24) at (2.5, 0) {};
		\node [style=dotsize] (25) at (3, 0) {};
		\node [style=dotsize] (26) at (3.5, 0) {};
		\node [style=bluegauge] (27) at (4, 0) {};
		\node [style=none] (29) at (-1, -0.5) {1};
		\node [style=none] (30) at (0, -0.5) {2};
		\node [style=none] (31) at (1, -0.5) {3};
		\node [style=none] (32) at (2, -0.5) {4};
		\node [style=none] (34) at (4, -0.5) {$2n$};
		\node [style=gauge3] (35) at (5.5, 2) {};
		\node [style=gauge3] (36) at (5.5, 1.25) {};
		\node [style=dotsize] (37) at (5.5, 0.5) {};
		\node [style=dotsize] (38) at (5.5, 0) {};
		\node [style=dotsize] (39) at (5.5, -0.5) {};
		\node [style=gauge3] (40) at (5.5, -1.25) {};
		\node [style=gauge3] (41) at (5.5, -2) {};
		\node [style=none] (42) at (6, 2) {1};
		\node [style=none] (43) at (6, 1.25) {1};
		\node [style=none] (44) at (6, -1.25) {1};
		\node [style=none] (45) at (6, -2) {1};
		\node [style=none] (46) at (6.5, 2) {};
		\node [style=none] (47) at (6.5, -2) {};
		\node [style=none] (48) at (7.4, 0) {$n-1$};
		\node [style=none] (49) at (-6.5, -2.25) {\scalebox{1.25}{Quiver for $\mathcal{N}_{\mathrm{USp}(2n)}$}};
		\node [style=none] (50) at (-2.5, 0.75) {\scalebox{1.25}{Explosion}};
		\node [style=none] (51) at (2, -2.25) {\scalebox{1.25}{Exploded quiver}};
		\node [style=none] (53) at (4, 2) {3};
		\node [style=miniU] (55) at (4, 1.5) {};
	\end{pgfonlayer}
	\begin{pgfonlayer}{edgelayer}
		\draw (0) to (3);
		\draw (3) to (1);
		\draw (1) to (4);
		\draw (8) to (9);
		\draw [style=->] (17.center) to (18.center);
		\draw (19) to (22);
		\draw (22) to (20);
		\draw (20) to (23);
		\draw (27) to (35);
		\draw (36) to (27);
		\draw (27) to (40);
		\draw (27) to (41);
		\draw [style=brace] (46.center) to (47.center);
		\draw (55) to (27);
	\end{pgfonlayer}
\end{tikzpicture}}
\end{adjustbox}
\end{equation}
However, the Coulomb branch for a $\mathrm{SO}(3)$ gauge group with $n$ flavors has trivial global symmetry (no contribution at order $t^2$ in the Hilbert series) for $n\geq 3$. Therefore, the global symmetry is $\mathrm{USp}(2n)\times \mathrm{U}(1)^{n-1}$ which is not the expected result for implosion. Based on the argument from global symmetry, we suggest the correct explosion is as in \eqref{implodeSO(2n+1)}.

\paragraph{Comment on bad quivers.} Since the quivers in this section are bad, and hence cannot be checked with current Hilbert series techniques, we should address what we know exactly about their moduli space. We will only consider the simplest example, the left quiver for $n=1$ in \eqref{implodeSO(2n+1)}. There is a choice to make, whether the leftmost gauge node is an $\mathrm{SO}(1)$ or $\mathrm{O}(1)$. The difference in moduli space is drastic. If we pick $\mathrm{SO}(1)$ then the theory is equivalent to $\mathrm{USp}(2)$ with 4 fundamental half hypermultiplets. The classical Higgs branch of this theory is the union of two cones, $
d_2=\mathbb{C}^2/\mathbb{Z}_2\cup\mathbb{C}^2/\mathbb{Z}_2$. It is the nilpotent cone of $\mathrm{O}(4)$. There is a $\mathbb{Z}_2=\mathrm{O}(1)$ symmetry exchanging the two cones. The Coulomb branch of this theory is $D_2=(\mathbb{R}^3\times S^1)/\mathbb{Z}_2$, where the $\mathbb{Z}_2$ acts on both the $\mathbb{R}^3$ and the $S^1$. The resulting space has two singular points, which are both of $A_1$ type \cite{Seiberg:1996nz}. The $\mathrm{O}(1)$ symmetry exchanges the two singularities. The classical Higgs branch is distorted quantum mechanically. In the full moduli space each $\mathbb{C}^2/\mathbb{Z}_2$ cone in the classical Higgs branch emanates from a different singularity of the Coulomb branch.

\begin{figure}
\centering
        \begin{tikzpicture}
            \node at (0,6) {\Large Quiver};
            \node at (6,6) {\Large Moduli Space Hasse diagram};
            \node at (0,4) {$\begin{tikzpicture}
                            \node[redgauge,label=below:{SO$(1)$}] (1) at (0,0) {};
                            \node[bluegauge,label=below:{$2$}] (2) at (1,0) {};
                            \node[flavourRed,label=below:{$3$}] (3) at (2,0) {};
                            \draw (1)--(2)--(3);
            \end{tikzpicture}$};
            \node[rotate=90] at (0.5,3.1) {$=$};
            \node at (0.9,2) {$\begin{tikzpicture}
                            \node[bluegauge,label=below:{$2$}] (2) at (1,0) {};
                            \node[flavourRed,label=below:{O$(4)$}] (3) at (2,0) {};
                            \draw (2)--(3);
            \end{tikzpicture}$};
            \node at (6,3) {$\begin{tikzpicture}
                            \node[dotsize] (01) at (-1,-2) {};
                            \node[dotsize] (03) at (1,-2) {};
                            \node[dotsize] (1) at (-2,0) {};
                            \node[dotsize] (2) at (0,0) {};
                            \node[dotsize] (3) at (2,0) {};
                            \draw[red] (01)--(2)--(03);
                            \draw[blue] (1)--(01) (03)--(3);
                            \draw[<->] (-1,-2.2) .. controls (0,-2.5) .. (1,-2.2);
                            \node at (0,-2.7) {O$(1)$};
                            \node at (-1.5,-1) {$A_1$};
                            \node at (-0.5,-1) {$A_1$};
                            \node at (0.5,-1) {$A_1$};
                            \node at (1.5,-1) {$A_1$};
            \end{tikzpicture}$};
            \node at (0.3,-1) {$\begin{tikzpicture}
                            \node[redgauge,label=below:{O$(1)$}] (1) at (0,0) {};
                            \node[bluegauge,label=below:{$2$}] (2) at (1,0) {};
                            \node[flavourRed,label=below:{SO$(3)$}] (3) at (2,0) {};
                            \draw (1)--(2)--(3);
            \end{tikzpicture}$};
            \node at (6,-1) {$\begin{tikzpicture}
                            \node[dotsize] (03) at (1,-2) {};
                            \node[dotsize] (2) at (0,0) {};
                            \node[dotsize] (3) at (2,0) {};
                            \draw[red] (2)--(03);
                            \draw[blue] (03)--(3);
                            \node at (0.5,-1) {$A_1$};
                            \node at (1.5,-1) {$A_1$};
            \end{tikzpicture}$};
        \end{tikzpicture}
        \caption{Depiction of the difference in moduli space choosing O$(1)$ over SO$(1)$ in the left quiver of \eqref{implodeSO(2n+1)} for $n=1$. In the Hasse diagram red lines denote Coulomb branch directions, while blue lines denote Higgs branch directions.}
        \label{fig:Z2}
\end{figure}
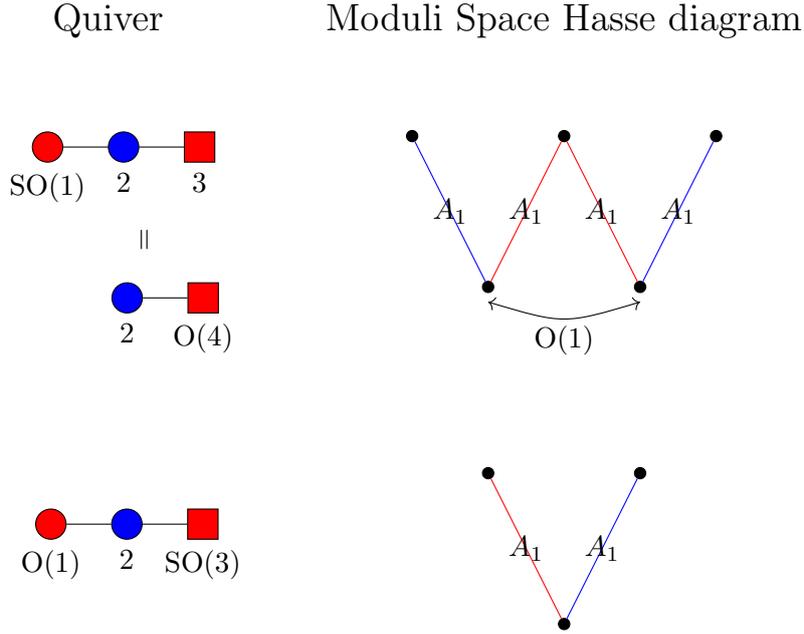

We can now gauge the $\mathrm{O}(1)$ symmetry, leading to the second choice in \eqref{implodeSO(2n+1)}, the leftmost node now being O rather than SO. The resulting moduli space is much simpler. The two cones in the Higgs branch are identified with each other, as are the two singularities in the Coulomb branch. The Coulomb branch of this quiver is expected to be $\mathbb{C}^2/\mathbb{Z}_2$ \cite{GaiottoWitten}. We can summarise this in Figure \ref{fig:Z2}, depicting the moduli spaces by their Hasse diagram \cite{Bourget:2019aer,Grimminger:2020dmg}, using red for Coulomb branch and blue for Higgs branch directions.

Since the Coulomb branch in the latter case is a hyper-K\"ahler cone, its Hilbert Series is computable. However the monopole formula does not give the desired result, but rather diverges. This is because something goes wrong with the conformal dimension formula for `bad' quivers. One could hope to adjust to this formula for certain cases, where the quiver is bad, but the Coulomb branch is a cone. We leave this for future work.

\appendix

\section{A bouquet of $\mathrm{USp}(2)$ nodes }
In this section, we list some interesting results for the explosion of $T[SO(2n+1)]$ quivers with the choice of a bouquet of $\mathrm{USp}(2)$ gauge nodes rather than $\mathrm{U}(1)$ gauge nodes that is discussed in Section \ref{implosions}. 

For a $\mathrm{USp}(2)$ gauge group with $n$ flavors, we need $n\geq 3$, otherwise the gauge node has negative imbalance and the monopole formula diverges. The first such case is $T[\mathrm{SO}(7)]$, where a bouquet of $\mathrm{USp}(2)$ gauge nodes takes the form:
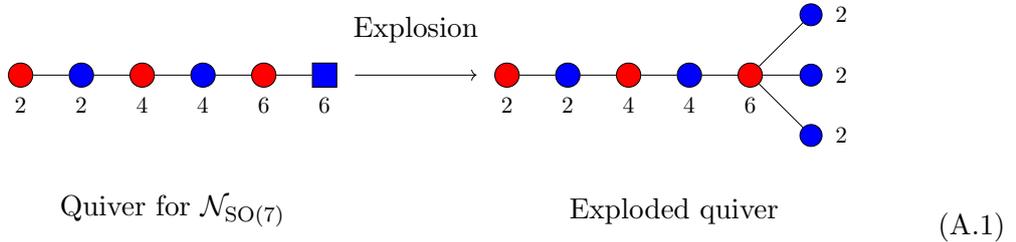
\begin{equation}
 \scalebox{.800}{\begin{tikzpicture}
	\begin{pgfonlayer}{nodelayer}
		\node [style=redgauge] (0) at (-9, 0) {};
		\node [style=redgauge] (1) at (-7, 0) {};
		\node [style=bluegauge] (3) at (-8, 0) {};
		\node [style=bluegauge] (4) at (-6, 0) {};
		\node [style=none] (10) at (-9, -0.5) {2};
		\node [style=none] (11) at (-8, -0.5) {2};
		\node [style=none] (12) at (-7, -0.5) {4};
		\node [style=none] (13) at (-6, -0.5) {4};
		\node [style=none] (15) at (-5, -0.5) {6};
		\node [style=none] (16) at (-4, -0.5) {6};
		\node [style=none] (17) at (-3.5, 0) {};
		\node [style=none] (18) at (-1.5, 0) {};
		\node [style=redgauge] (19) at (-1, 0) {};
		\node [style=redgauge] (20) at (1, 0) {};
		\node [style=bluegauge] (22) at (0, 0) {};
		\node [style=bluegauge] (23) at (2, 0) {};
		\node [style=none] (29) at (-1, -0.5) {2};
		\node [style=none] (30) at (0, -0.5) {2};
		\node [style=none] (31) at (1, -0.5) {4};
		\node [style=none] (32) at (2, -0.5) {4};
		\node [style=none] (34) at (3, -0.5) {6};
		\node [style=none] (49) at (-6.5, -2.25) {\scalebox{1.25}{Quiver for $\mathcal{N}_{\mathrm{SO}(7)}$}};
		\node [style=none] (50) at (-2.5, 0.75) {\scalebox{1.25}{Explosion}};
		\node [style=none] (51) at (1.75, -2.25) {\scalebox{1.25}{Exploded quiver}};
		\node [style=flavourBlue] (52) at (-4, 0) {};
		\node [style=redgauge] (53) at (-5, 0) {};
		\node [style=redgauge] (54) at (3, 0) {};
		\node [style=miniBlue] (55) at (4, 1) {};
		\node [style=miniBlue] (56) at (4, 0) {};
		\node [style=miniBlue] (57) at (4, -1) {};
		\node [style=none] (58) at (4.5, 1) {2};
		\node [style=none] (59) at (4.5, 0) {2};
		\node [style=none] (60) at (4.5, -1) {2};
	\end{pgfonlayer}
	\begin{pgfonlayer}{edgelayer}
		\draw (0) to (3);
		\draw (3) to (1);
		\draw (1) to (4);
		\draw [style=->] (17.center) to (18.center);
		\draw (19) to (22);
		\draw (22) to (20);
		\draw (20) to (23);
		\draw (53) to (52);
		\draw (54) to (55);
		\draw (54) to (56);
		\draw (54) to (57);
		\draw (23) to (54);
		\draw (4) to (53);
	\end{pgfonlayer}
\end{tikzpicture}}
\label{wrongimplosion}
\end{equation}
We see here that all the gauge nodes are balanced, yet the Coulomb branch Hilbert series diverges. This is not too surprising as a quiver can be bad even if none of the individual gauge nodes are bad. This is often observed for orthosymplectic quivers. 

For $n\geq 4$, the Coulomb branch Hilbert series no longer diverges. Using the fact that the Coulomb branch of $\mathrm{USp}(2)$ gauge group with $n$ flavors and $n\geq 3$ has trivial global symmetry, we know that the global symmetry of the exploded quiver will just be $\mathrm{SO}(2n+1)$. This is reported in Table \ref{C1implosion}. However, for $n=4$, there is a contribution from $\mathrm{HS}_{\mathbb{Z}+\frac{1}{2}}$   at order $t^2$, hence enhancing the global symmetry. The enhancement from $\mathrm{HS}_{\mathbb{Z}+\frac{1}{2}}$ comes in the spinor representation of $\mathrm{SO}(9)$ which results in $F_4$ of dimension 52.

 \afterpage{\begin{landscape}
\begin{table}[!ht]
\begin{adjustbox}{center}
\small	\begin{tabular}{|c|c|c|c|c|}
\hline
Orbit &	 Quiver & \begin{tabular}{c}Global \\ Symmetry \end{tabular}& Hilbert Series  & \begin{tabular}{c} Plethystic \\ Logarithm \end{tabular} \\ 
\hline
$\mathfrak{so}(9)$ &
\begin{tabular}{c}
\centering
 \scalebox{0.75}{        
    \begin{tikzpicture}
				\tikzstyle{gauge1} = [draw=none,minimum size=0.35cm,fill=blue,circle, draw];
				\tikzstyle{gauge2} = [draw=none,minimum size=0.35cm,fill=red,circle, draw];
            	\tikzstyle{flavour1}  = [draw=none,minimum size=0.35cm,fill=blue,circle, draw];
				\tikzstyle{flavour2} = [draw=none,minimum size=0.35cm,fill=red, regular polygon,regular polygon sides=4,draw];
				\node (g1) [gauge2,label=below:{$2$}] {};
				\node (g2) [gauge1,right of=g1,label=below:{$2$}] {};
				\node (g3) [gauge2,right of=g2,label=below:{$4$}] {};
				\node (g4) [gauge1,right of=g3,label=below:{$4$}] {};
				\node (g5) [gauge2,right of=g4,label=below:{$6$}] {};
				\node (g6) [gauge1,right of=g5,label=below:{$6$}] {};
				\node (g7) [gauge2,right of=g6,label=below:{$8$}] {};
				\node (f2) [flavour1,right of =g7,above of=g7, label=above:{$2$}] {};
				\node (f1) [flavour1,right of=f2,label=above:{$2$}] {};                
				\node (f3) [flavour1,right of =g7,below of=g7,label=below:{$2$}] {};
				\node (f4) [flavour1,right of=f3,label=below:{$2$}] {};
				\draw (g7)--(f1) (g7)--(f2) (g7)--(f3) (g7)--(f4) 
				(g1)--(g2) (g2)--(g3) (g3)--(g4) (g4)--(g5) (g5)--(g6)  (g6)--(g7);
				\end{tikzpicture}  }
\end{tabular}
& 		
\centering	
	$F_4$ &
\begin{tabular}{l}
      \parbox[t]{6cm}{
  $1 + 52 t^2 + 1455 t^4 + 28834 t^6 + 449122 t^8 + 5793780 t^{10} + 63853945 t^{12} + 613989328 t^{14} + 5232181818 t^{16} +  40010832518 t^{18} + 277431116267 t^{20}+ O\left(t^{22}\right)$  }\\
       \end{tabular}
       &
       \begin{tabular}{c}
       \parbox{2.5cm}{$   52 t^2 + 77 t^4 + 26 t^6 - 2394 t^8 - 5442 t^{10} +O\left(t^{12}\right)   $} 
       \end{tabular}  \\  
\hline
       $\mathfrak{so}(11)$&
\begin{tabular}{c}	
\centering
 \scalebox{0.6}{
         \begin{tikzpicture}
				\tikzstyle{gauge1} = [draw=none,minimum size=0.35cm,fill=blue,circle, draw];
				\tikzstyle{gauge2} = [draw=none,minimum size=0.35cm,fill=red,circle, draw];
            	\tikzstyle{flavour1}  = [draw=none,minimum size=0.35cm,fill=blue,circle, draw];
				\tikzstyle{flavour2} = [draw=none,minimum size=0.35cm,fill=red, regular polygon,regular polygon sides=4,draw];
				\node (g1) [gauge2,label=below:{$2$}] {};
				\node (g2) [gauge1,right of=g1,label=below:{$2$}] {};
				\node (g3) [gauge2,right of=g2,label=below:{$4$}] {};
				\node (g4) [gauge1,right of=g3,label=below:{$4$}] {};
				\node (g5) [gauge2,right of=g4,label=below:{$6$}] {};
				\node (g6) [gauge1,right of=g5,label=below:{$6$}] {};
				\node (g7) [gauge2,right of=g6,label=below:{$8$}] {};
				\node (g8) [gauge1,right of=g7,label=below:{$8$}] {};
				\node (g9) [gauge2,right of=g8,label=below:{$10$}] {};
				\node (f2) [flavour1,right of =g9,above of=g9, label=above:{$2$}] {};
				\node (f1) [flavour1,right of=f2,label=above:{$2$}] {};                
				\node (f3) [flavour1,right of =g9,below of=g9,label=below:{$2$}] {};
				\node (f4) [flavour1,right of=f3,label=below:{$2$}] {};
			\node (f5) [flavour1,right of=g9,label=right:{$2$}] {};	
				\draw (g9)--(f5) (g9)--(f1) (g9)--(f2) (g9)--(f3) (g9)--(f4) 
				(g1)--(g2) (g2)--(g3) (g3)--(g4) (g4)--(g5) (g5)--(g6)  (g6)--(g7) (g7)--(g8) (g8)--(g9);
				\end{tikzpicture}  }
\end{tabular}
	& 		$\mathrm{SO}(11)$ &	
\begin{tabular}{c}  
\parbox[t]{6cm}{$1 + 55 t^2 + 1544 t^4 + 32 t^5 + 29535 t^6 + 1888 t^7 + 433464 t^8 + 56608 t^9 + 5209798 t^{10} + 1151360 t^{11} + 53454368 t^{12} + 17885120 t^{13} + 482022542 t^{14} + 226457760 t^{15} + 3904141695 t^{16} +  2435618944 t^{17} + 28892424245 t^{18} + 22895407232 t^{19}++198088208252 t^{20}+192087708320 t^{21}+O\left(t^{22}\right)$}\end{tabular}
	&
\begin{tabular}{c}
	\parbox[t]{2.5cm}
	{$55 t^2 + 4 t^4 + 32 t^5 + 55 t^6 + 128 t^7 - t^8 + 160 t^9 + 429 t^{10} +O\left(t^{11}\right)$}
	\end{tabular} \\
\hline
 	 $ \mathfrak{so}(13)$ &   
\begin{tabular}{c}	
\centering
\scalebox{0.5}{
\begin{tikzpicture}
				\tikzstyle{gauge1} = [draw=none,minimum size=0.35cm,fill=blue,circle, draw];
				\tikzstyle{gauge2} = [draw=none,minimum size=0.35cm,fill=red,circle, draw];
            	\tikzstyle{flavour1}  = [draw=none,minimum size=0.35cm,fill=blue,circle, draw];
				\tikzstyle{flavour2} = [draw=none,minimum size=0.35cm,fill=red, regular polygon,regular polygon sides=4,draw];
				\node (g1) [gauge2,label=below:{$2$}] {};
				\node (g2) [gauge1,right of=g1,label=below:{$2$}] {};
				\node (g3) [gauge2,right of=g2,label=below:{$4$}] {};
				\node (g4) [gauge1,right of=g3,label=below:{$4$}] {};
				\node (g5) [gauge2,right of=g4,label=below:{$6$}] {};
				\node (g6) [gauge1,right of=g5,label=below:{$6$}] {};
				\node (g7) [gauge2,right of=g6,label=below:{$8$}] {};
				\node (g8) [gauge1,right of=g7,label=below:{$8$}] {};
				\node (g9) [gauge2,right of=g8,label=below:{$10$}] {};
		    	\node (g10) [gauge1,right of=g9,label=below:{$10$}] {};	
		    	\node (g11) [gauge2,right of=g10,label=below:{$12$}] {};	
				\node (f2) [flavour1,right of =g11,above of=g11, label=above:{$2$}] {};
				\node (f1) [flavour1,right of=f2,label=above:{$2$}] {};                
				\node (f3) [flavour1,right of =g11,below of=g11,label=below:{$2$}] {};
				\node (f4) [flavour1,right of=f3,label=below:{$2$}] {};
			\node (f5) [flavour1,right of=f4,label=below:{$2$}] {};
				\node (f6) [flavour1,right of=f1,label=above:{$2$}] {};
				\draw (g11)--(f5) (g11)--(f1) (g11)--(f2) (g11)--(f3) (g11)--(f4) 
				(g1)--(g2) (g2)--(g3) (g3)--(g4) (g4)--(g5) (g5)--(g6)  (g6)--(g7) (g7)--(g8) (g8)--(g9) (g9)--(g10) (g10)--(g11) (g11)--(f6);
				\end{tikzpicture}  }
\end{tabular}
	& 			$\mathrm{SO}(13)$ &
\begin{tabular}{c}
\parbox[t]{6cm}
{$ 1 + 78 t^2 + 3086 t^4 + 82550 t^6 + 1679237 t^8 +64t^9 +27703312 t^{10}+5312 t^{11} + 386049641 t^{12} +223040 t^{13}+ 4673378944 t^{14}+6315904 t^{15} + 50165955517 t^{16}+135683648 t^{17} + 485045344353 t^{18 }+2358618624 t^{19}+ 4276770597522 t^{20 }+ 34556791744 t^{21}+ O\left(t^{22}\right)$}
\end{tabular}
&
\begin{tabular}{c}
\parbox[t]{2.5cm}{$78 t^2 + 5 t^4 + 77 t^8 + 64 t^9 + O\left(t^{11}\right)$}
\end{tabular} \\ 
\hline
	\end{tabular}
\end{adjustbox}
\caption{Quivers whose Coulomb branches are the nilpotent cones of $\mathrm{SO}(2n+1)$ for $n=4,5,6$ are fully `exploded'. The $\mathrm{USp}(2n)$ flavour node turns into $n$ $\mathrm{USp}(2)$ gauge nodes.  The Hilbert series and the plethystic logarithm are provided.}
\label{C1implosion}
\end{table}
\end{landscape}}

\clearpage

\section*{Acknowledgements}
We would like to thank Rudolph Kalveks for helpful discussions. The work of AB, JFG, AH and ZZ is supported by STFC grants ST/P000762/1 and ST/T000791/1.

\bibliographystyle{JHEP}
\bibliography{bibli.bib}

\end{document}